# Intermediate Points for Missions to Interstellar Objects Using Optimum Interplanetary Trajectory Software

Author: Adam Hibberd[a]
Affiliation: [a] Initiative for Interstellar Studies (i4is).
Address: 27/29 South Lambeth Road London, SW8 1SZ United Kingdom
Email: adam.hibberd@ntlworld.com

## ABSTRACT

This paper explicates the concept of an 'Intermediate Point' (IP), its incorporation as a node along an interplanetary trajectory, and how it permits the determination and optimization of trajectories to interstellar objects (ISOs). IPs can be used to model Solar Oberth Manoeuvres (SOM) as well as $V_\infty$ Leveraging Manoeuvres (VLM). The SOM has been established theoretically as an important mechanism by which rapid exit from the solar system and intercept of ISOs can both be achieved; the VLM has been demonstrated practically as a method of reducing overall mission $\Delta V$ as well as the Characteristic Energy, $C_3$, at launch. Thus via these two applications, the feasibility of missions to interstellar objects (ISOs) such as 1I/'Oumuamua can be analysed. The interplanetary trajectory optimizer tool exploited for this analysis, OITS, permits IP selection as an encounter option along the interplanetary trajectory in question. OITS adopts the assumption of impulsive thrust at discrete points along the trajectory, an assumption which is generally valid for high thrust propulsion scenarios, like chemical or nuclear thermal for instance.

*Keywords*: Interstellar Objects, Intermediate Points, Trajectory Optimization, Solar Oberth Manoeuvres, $V_\infty$ Leveraging Manoeuvres

**Intermediate Points for Missions to Interstellar Objects Using Optimum Interplanetary Trajectory Software**

**1.) Introduction**

For humanity the realisation of travel to the stars beyond our own is confounded by a combination of their enormous distances and the theoretical constraint on the speed of the spacecraft, the speed of light, c. Thus the nearest star to our own, Proxima Centauri at 4.2 light years, would take tens of thousands of years using present day chemical propulsion technology.

However the arrival and discovery of interstellar objects (ISOs), interlopers to our solar system originating from elsewhere in our galaxy, has conveniently saved us the trouble of such prolonged journey times – material from other systems has arrived on our door-step for us to study if we so desire.

Since 1I/'Oumuamua [1], was discovered on an unquestionably hyperbolic orbit, i.e. with e = 1.2 ( > 1), the possibility of flyby or rendezvous missions have been the subject of speculation and debate. With their significant heliocentric hyperbolic excess speeds (for 'Oumuamua $V_\infty > 26.3 kms^{-1}$), the problem of the feasibility of spacecraft missions to ISOs after they have passed beyond their perihelia and are leaving the solar system, has been debated by scientists [2,3]. The resolution of this problem is no longer a matter for conjecture, it can be quantified by the application of the necessary software tool, for example Optimum Interplanetary Trajectory Software (OITS) [4]. OITS was originally developed for the express purpose of investigation of missions to objects belonging to our own solar system, hence 'interplanetary'. Upon the discovery of 'Oumuamua with all its unique characteristics, OITS was adapted with some minor alterations to allow exploration of intercept trajectories to this ISO [5,6,7,8]. With the subsequent detection in 2019 of the second ISO, 2I/Borisov, further research was conducted into missions to this object [9].

The key facility within OITS allowing investigations of this kind is the 'Intermediate Point' (IP), which can be chosen by the user as an alternative to a planetary encounter and enables the user to incorporate what is known as a *'Solar Oberth Manoeuvre'* (SOM) along the interplanetary trajectory. To elucidate, a SOM is a mechanism by which a high exit speed out of the solar system can be generated whilst utilizing a minimum magnitude ΔV applied by the spacecraft propulsion system [10,11].

A further application for the IP, not necessarily confined to ISO missions, is the modelling of '$V_\infty$ *Leveraging Manoeuvres'* (VLM) which reduces the total ΔV needed to arrive at the target which might be an outer planet, for example [12].

## 2.) Basic Theory

We have *Nbody* celestial bodies in sequence, $P_1, P_2, P_3...P_{Nbody}$. We can associate with these bodies *Nbody* times, $t_1, t_2, t_3...t_{Nbody}$. By using the NASA Horizons service in conjunction with the NASA SPICE toolkit, extremely accurate positions, $r_i$ $i=1...Nbody$, for all these objects can be derived as a function of time. If we assume the sun is the only gravitating body acting on the spacecraft, then for two bodies, let's say $P_1$ and $P_2$ at times $t_1$ and $t_2$ respectively, the Lambert problem [13] can be solved to allow computation of two interplanetary trajectories from $P_1$ to $P_2$, short way, '*sw*', and long way, '*lw*', with the necessary endpoint times and positions being respected.

**2.1) Universal Variable Formulation**

The technique employed is the 'Universal Variable Formulation' [13]. OITS does not seek solutions involving multiple orbital revolutions, which restricts the solution to two possible transfer orbits. The equations are provided below:

$$\Delta\theta = \text{acos}\left(\frac{r_1 \cdot r_2}{r_1\, r_2}\right) \tag{1}$$

$$\Delta v = \Delta\theta\,,\; \Delta v = 2\pi - \Delta\theta \tag{2}$$

$$A = \frac{\sqrt{r_1 r_2}\sin\Delta v}{\sqrt{1-\cos\Delta v}} \tag{3}$$

$$y = r_1 + r_2 - A\frac{(1-zS)}{\sqrt{C}} \tag{4}$$

$$x = \sqrt{\frac{y}{C}} \tag{5}$$

$$t = \frac{x^3 S + A\sqrt{y}}{\sqrt{\mu}} \tag{6}$$

Where:

$$C = C(z) = \frac{1-\cos\sqrt{z}}{z} \tag{7}$$

$$S = S(z) = \frac{\sqrt{z}-\sin\sqrt{z}}{\sqrt{z^3}} \tag{8}$$

$$z < (2\pi)^2 \tag{9}$$

Note that μ is the gravitational mass of the sun. The above three equations (4)-(6) are solved for *x*, *y*, *z*, where $t = t_2 - t_1$, $r_1$, $r_2$ are the knowns (*Δv*, *Δθ*, *A* being functions of these) and *x*, *y*, *z* are the unknowns.

Having solved for *x*, *y*, *z*, the following equations allow the heliocentric departure velocity at $P_1$, **VD** and arrival velocity at $P_2$ **VA** of the interplanetary trajectory to be calculated:

$$f = 1 - \frac{y}{r_1} \tag{10}$$

$$g = A\sqrt{\frac{y}{\mu}} \tag{11}$$

$$\dot{g} = 1 - \frac{y}{r_2} \tag{12}$$

$$\boldsymbol{VD} = \frac{r_2 - f r_1}{g} \tag{13}$$

$$\boldsymbol{VA} = \frac{\dot{g} r_2 - r_1}{g} \tag{14}$$

This process can be repeated for all trajectories *(sw$_j$, lw$_j$) j=1…Nbody-1*.

### 2.2) Solution for Two Bodies – Home and Target

With *Nbody=2*, and the addition of an equation to give the gradient *dt/dz,* the above can be solved using a Newton iteration. Two *ΔVs* can then be computed depending on whether the trajectory is along the short way (*Δv = Δθ*) or long way (*Δv = 2π - Δθ*) Furthermore for both routes there will be two possible alternative values of *ΔV*, depending on whether it is a flyby mission or a rendezvous mission:

$$\Delta V_k = |\boldsymbol{VD}_k - \boldsymbol{V}_1| \qquad \text{flyby} \tag{15}$$

$$\Delta V_k = |\boldsymbol{VD}_k - \boldsymbol{V}_1| + |\boldsymbol{VA}_k - \boldsymbol{V}_2| \quad \text{rendezvous} \tag{16}$$

Where k =1 or 2 so that:

$$\Delta v = \Delta\theta, \qquad k = 1 \tag{17}$$

$$\Delta v = 2\pi - \Delta\theta, \quad k = 2 \tag{18}$$

Note *V$_1$* and *V$_2$* are the velocities of the home and arrival celestial bodies respectively.

## 2.3) Solutions for *Nbody > 2*

For *Nbody>2* bodies then there are *Ntrans=Nbody-1* connecting interplanetary transfer trajectories, each with a *lw* and a *sw*, corresponding to $N=2^{Ntrans}$ permutations. Each of these *Ntrans* transfer trajectories can be solved in the same manner explained in Section 2.2. As an example let us assume *Nbody=4* and therefore there are two terminal planets $P_1$ & $P_4$ and also $NP=2^3=8$ possible permutations of trajectory, connecting $P_1$, $P_2$, $P_3$ & $P_4$ at times $t_1$, $t_2$, $t_3$ & $t_4$. The situation is illustrated schematically in Figure 1.

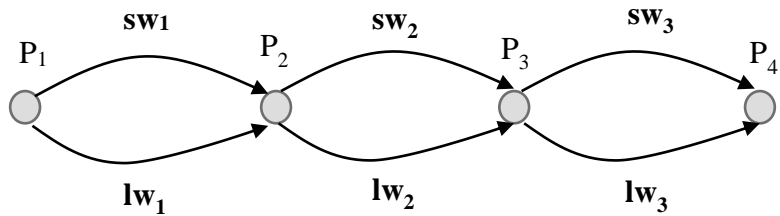

*Figure 1 Schematic of Interplanetary Trajectory with Nbody=4 Planets*

If we zoom in on one of the non-terminal celestial bodies, $P_3$ for example, then we are presented with Figure 2.

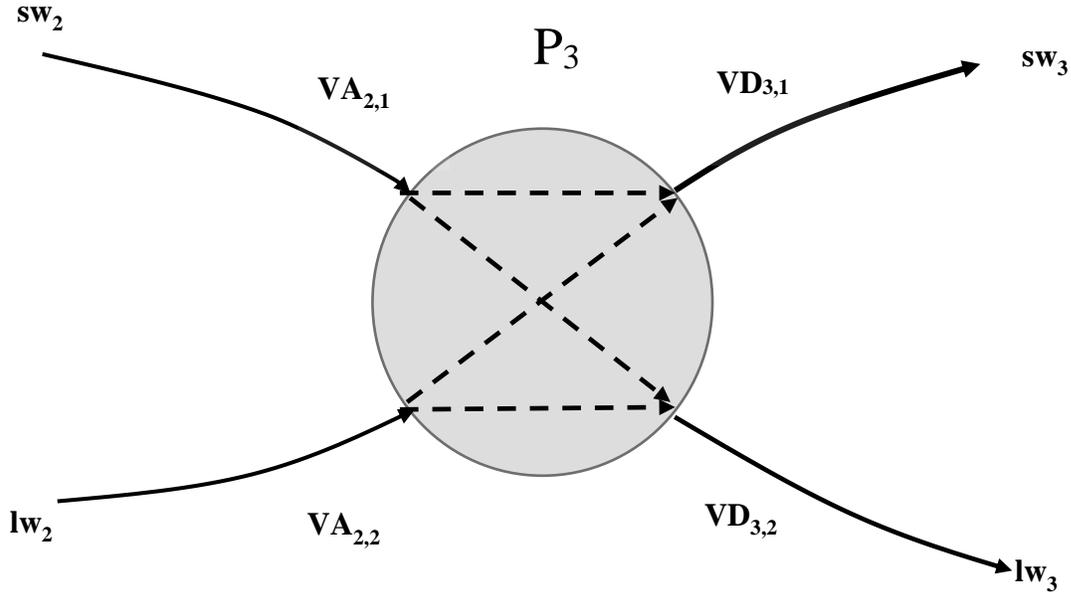

*Figure 2 Encounter Possiblities for Non-Terminal Planetary Encounter*

Focusing on a non-terminal Planet $P_i$, we can see there are four possible alternative encounter trajectories that the spacecraft may travel along as it encounters the planet. Let us take one possible arrival and departure velocity, given by

$$VA_{i-1,k1} \text{ \& } VD_{i,k2}$$

where $2 \leq i \leq Ntrans$

$k1 = 1 \text{ or } 2 \text{ and } k2 = 1 \text{ or } 2.$

Furthermore the angle between them is given by:

$$cos\alpha = \frac{VA_{i-1,k1} \cdot VD_{i,k2}}{|VA_{i-1,k1}| |VD_{i,k2}|} \tag{19}$$

We now make the following assumptions:

1) that the hyperbolic excess velocities at arrival and departure of $P_i$ are $VA_{i-1,k1}$ and $VD_{i,k2}$ respectively

2) that the encounter trajectory of the spacecraft w.r.t. $P_i$ is entirely influenced by the gravitational attraction of $P_i$, and we ignore the influences of any other bodies, such as the sun (i.e. patched conic).

3) a velocity increment of magnitude $\Delta V$ is applied in the plane containing **$VA_{i-1,k1}$** and **$VD_{i,k2}$**

4) further that this $\Delta V$ is applied at the periapsis point w.r.t. $P_i$

5) This $\Delta V$ is directed tangential to the trajectory

The above assumptions require two hyperbolas, Hyp and Hyp' as shown in Figure 3 connected at the periapsis point. The situation can be encapsulated in the three simultaneous equations:

$$\cos\gamma = -\frac{\mu_i}{(\mu_i + r_{per} VA_{i-1,k1}^2)} \tag{20}$$

$$\cos\gamma' = -\frac{\mu_i}{(\mu_i + r_{per} VD_{i,k2}^2)} \tag{21}$$

$$\alpha + \pi = \gamma + \gamma' \tag{22}$$

Where $\gamma$ is the angle subtended at the centre of $P_i$ between the direction of the hyperbolic arrival asymptote and the periapsis point; and $\gamma'$ is the angle subtended at the centre of $P_i$ between the periapsis point and hyperbolic escape asymptote. In addition $r_{per}$ is the common periapsis where Hyp and Hyp' connect and $\mu_i$ is the gravitational mass of $P_i$. The three simultaneous equations (20)-(22) have three unknowns $\gamma, \gamma', r_{per}$. It can be shown that, solving for $\gamma$ is a matter of finding the solution to the following equation:

$$\left((VD_{i,k2}^2 - VA_{i-1,k1}^2)\cos\gamma + VD_{i,k2}^2\right)\cos(\alpha - \gamma) + VA_{i-1,k1}^2 \cos\gamma = 0 \tag{23}$$

A derivative of the function on the left hand side of (23) can be found enabling application of a Newton iteration to solve this equation. Having converged to the solution $\gamma$, then $r_{per}$ and $\Delta V$ can be computed. Furthermore inclination, $i$, argument of periapsis $\omega$, and longitude of ascending node, $\Omega$, (all w.r.t. the Planet $P_i$) can be determined using the plane containing $\mathbf{VA}_{i-1,k1}$ and $\mathbf{VD}_{i,k2}$.

Though the precise equations (20)-(23) are different to those in [14,15], nevertheless the assumptions employed by this algorithm and the solutions found will be identical to those applied.

This procedure outlined above can be performed on each non-terminal planet and there will be in total $2^{Ntrans}$ permutations, each with a different combination of $\Delta Vs$ and each with a different sum-total. Of all these sum-totals, the minimum is chosen.

### 2.4) General Formulation of Problem

The general mathematical statement of the problem as follows.

Given *Nbody* planets $P_i$, $i=1...Nbody$, each associated with a time $t_i$, $i=1...Nbody$, we can generate the total $\Delta V$ (which will be the minimum of a total of $2^{Ntrans}$ permutations) as a function of these times, as follows:

$$\Delta V = \Phi(t_1, t_2, .... t_{Nbody}) \qquad (24)$$

The algorithm to generate $\Phi$ is elucidated in Sections 2.1 to 2.3.

Thus the problem is to minimize $\Phi$ whilst satisfying the following constraints:

a) The periapsis at each of the non-terminal encounters should be greater than a user-specifed distance **RMIN'**, thus

$$\mathbf{RPER}_i > \mathbf{RMIN}'_i \qquad i=2...Ntrans \qquad (25)$$

b) Where applicable, the perihelion along the trajectory arc from $i=1..Ntrans$ must be greater than a user-specified distance **RPERIH'**, thus

$$\mathbf{RPERIH}_i \geq \mathbf{RPERIH}'_i \qquad i=1...Ntrans \qquad (26)$$

The solution of finding the minimum of (24) subject to the constraints (25) & (26) amounts to a Non-Linear Programming (NLP) problem and for OITS two NLP tools were selected for solving this task, NOMAD [16] and MIDACO [17].

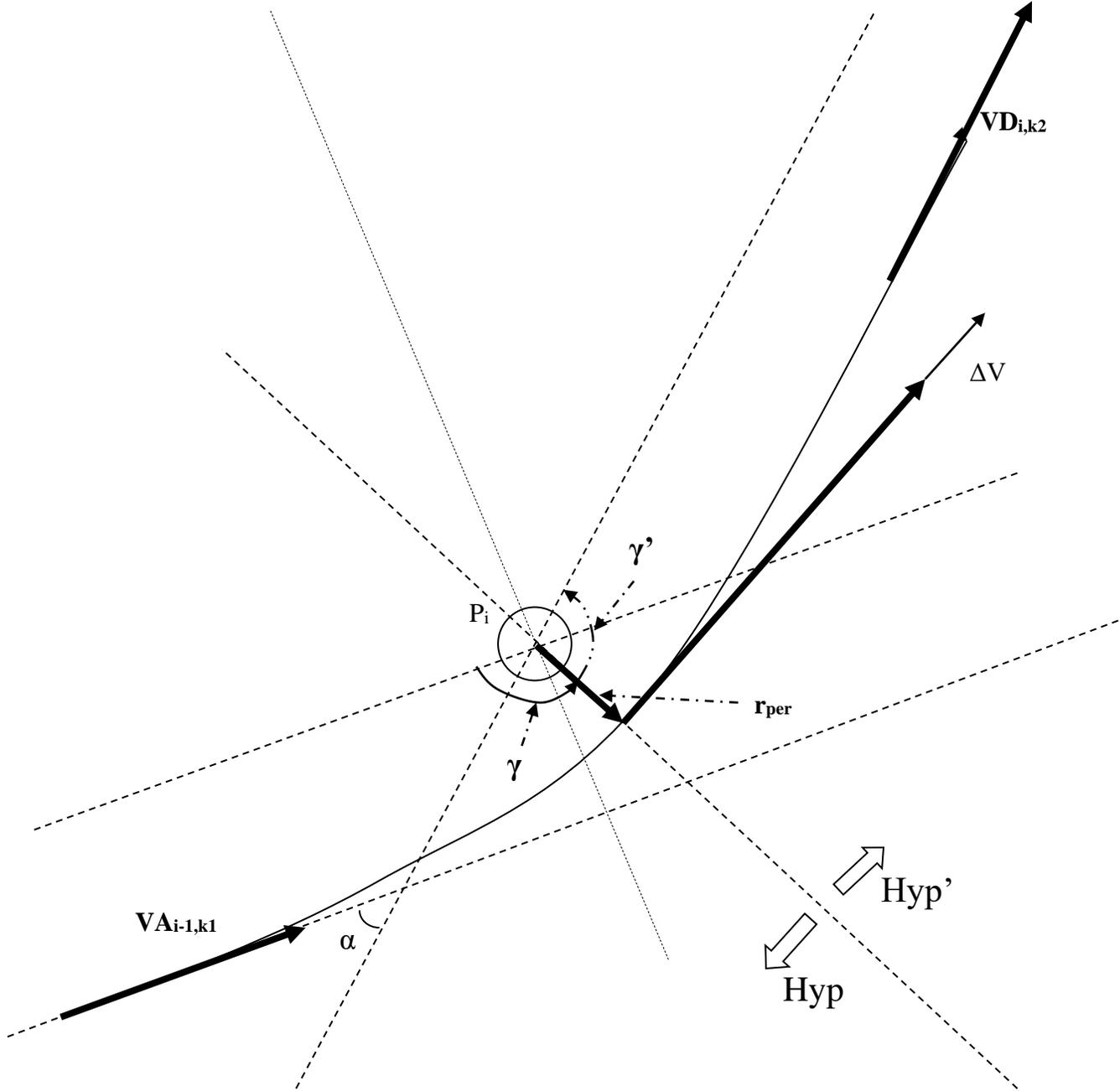

*Figure 3 Encounter with a Planet*

## 3) Intermediate Points

### 3.1) History

With the arrival of the first definite interstellar object (ISO) 1I/'Oumuamua in our solar system, the question arose as to whether the algorithm used by the OITS software (as expounded in Section 2) could be exploited to study the feasibility of missions. Under the Project Lyra initiative, inaugurated by the Initiative for Interstellar Studies (i4is), the requirement for a 'Solar Oberth' Manoeuvre (SOM) [10] had already been identified. The reasoning behind this choice of a SOM was two-fold. Firstly 'Oumuamua had been discovered significantly after perihelion, receding from the sun with considerable speed, necessitating a 'catch-up' approach. (The heliocentric hyperbolic excess speed of 'Oumuamua is upwards of 26 kms$^{-1}$, thus requiring a spacecraft speed of at least this value.) Secondly, with an extra layer of complexity, the inclination of 'Oumuamua's orbit is relatively high, resulting in a significant displacement from the ecliptic plane at intercept, a problem which nevertheless yields readily to the application of a SOM. However the OITS algorithm in its form at that time was insufficient for modelling a SOM, because encounter events along the trajectory were restricted to celestial bodies and corresponding binary SPICE kernel files retrieved from the NASA Horizons service. To overcome this limitation the notion of an 'Intermediate Point' was conceived and then applied as an additional functionality to the software.

## 3.2) Definition

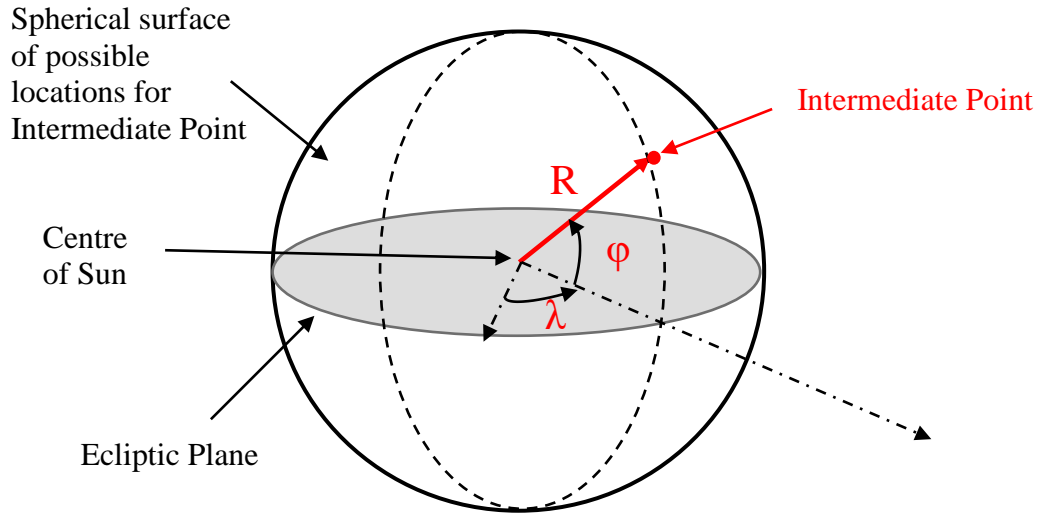

**Intermediate Point (IP) Definition**

**R = User-Specified Radius**

**λ = Heliocentric Longitude to be Optimized**

**φ = Heliocentric Latitude to be Optimized**

*Figure 4 Definition of Intermediate Point*

An Intermediate Point (IP) is a user-specified distance from the centre of the solar system. This can be selected instead of a celestial body as an encounter option in the user-specified sequence. Figure 4 is a diagram showing the definition of an IP.

The values of the polar angles $\lambda$ and $\varphi$ for the IP under consideration are included with the times $t_i$, $i=1...Nbody$ as parameters to be optimized by the NLP tool applied (refer Section 2.4). Thus for one IP along the trajectory equation 24 can be rewritten:

$$\Delta V = \Phi(t_1, t_2, ....t_{Nbody}, \lambda, \varphi) \qquad (27)$$

It is now a question of finding the minimum of function Φ, subject to user-specified limits on $t_i$ and also $-\pi \leq \lambda \leq \pi$ and $-\pi/2 \leq \varphi \leq \pi/2$.

### 3.3) Using an Intermediate Point to Model a Solar Oberth Manoeuvre

For a typical mission scenario to 1I/'Oumuamua, the sequence E-J-IP-1I might be applied. This abbreviation indicates a launch from Earth (E), an encounter with Jupiter (J), an Intermediate Point (IP), and finally arrival at 'Oumuamua (1I). The distance from the sun's centre for the IP might then be chosen as 0.01395 au (equivalent to 3 Solar Radii). In addition, two minimum perihelia constraints are also necessary. The situation is illustrated in Figure 5 & Figure 6.

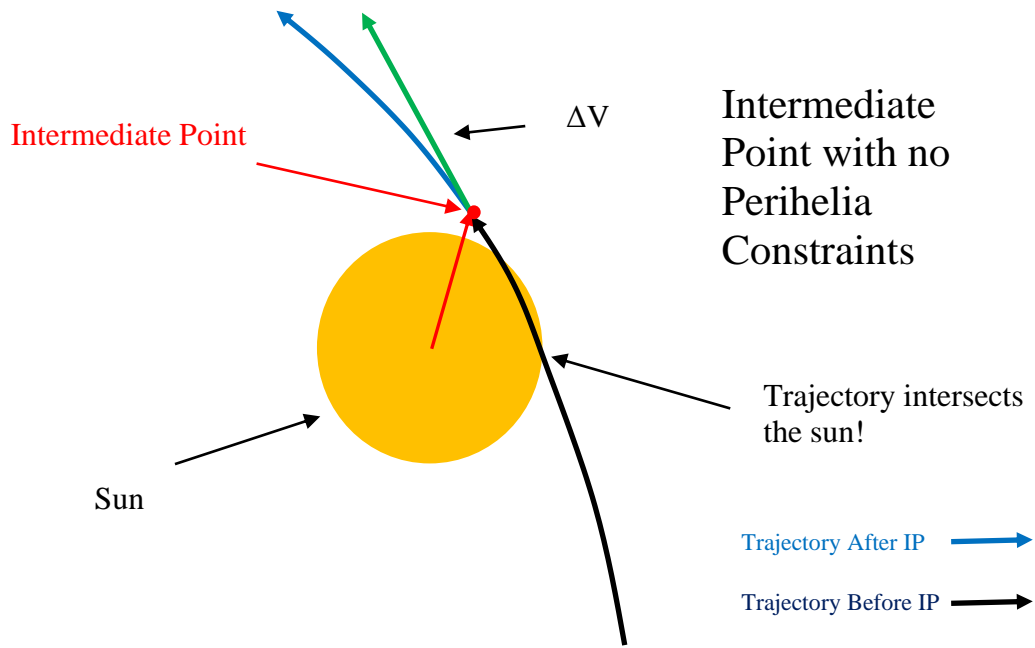

*Figure 5 Intermediate Point for SOM, WITHOUT Perihelia Constraints*

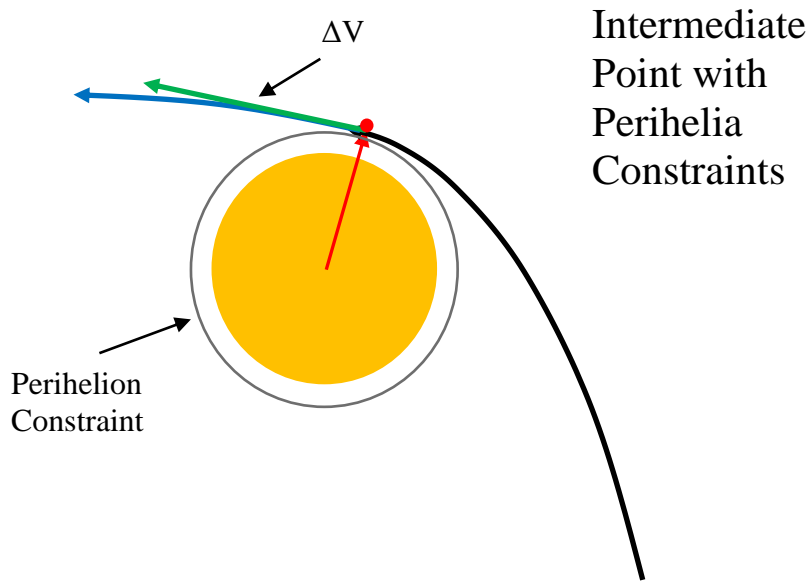

*Figure 6 Intermediate Point for SOM, WITH Perihelia Constraints*

In Figure 6 we see the result of application of a minimum perihelia constraint on both the trajectory leading to the IP and that leaving the IP. This compels the IP to be a perihelion point, as required of a SOM.

**3.4) Using an Intermediate Point to Model a $V_\infty$ Leveraging Manoeuvre**

For an interplanetary mission, a VLM [12] is a means by which the hyperbolic excess speed ($V_\infty$) with respect to a planet P can be amplified by conducting a Deep Space Manoeuvre (DSM) and then returning to the planet P for a gravity assist. By so doing, a considerable reduction in *ΔV* along the trajectory can be achieved enabling access to the outer planets, Jupiter for example, with less fuel mass. The return to P is accomplished after approximately, *but not exactly*, a time lapse of *n* cycles of P, i.e. *nT*. For example if the planet P is Earth, then the time elapsed will be approximately *n * 365.25* days where *n = 1, 2, or 3* etc.

It is emphasized that this elapsed time must be either side of *nT*, as theoretically there is no benefit from the situation where it is exactly *nT*. To elaborate, there are two alternative scenarios, i.e. *n-* and *n+*, the former when the return to P is slightly sooner than *nT* and the latter when it is slightly later than *nT*. Thus the interplanetary trajectory which can be abbreviated to P-DSM-P can be considered as *resonant* with the orbit of P.

A VLM can be modelled by incorporating an IP in between two successive encounters of the planet in question. The value of *R* specified for the IP can be derived as follows. Let us consider a Leveraging at a resonance *n* with the orbit of P, this planet having a time period $T_P$. By Kepler's third law and Newton's law of gravitation, we have for the (approximately) elliptical leg of the journey P-DSM-P:

$$nT_p = 2\pi \cdot \sqrt{\frac{a^3}{\mu}} \qquad (28)$$

Where a is the semi-major axis of the interplanetary trajectory P-DSM-P.

Rearranging we have:

$$a = \sqrt[3]{\mu \left(\frac{nT_p}{2\pi}\right)^2} \qquad (29)$$

Now along P-DSM-P, the spacecraft perihelion will be at $R_P$ and the aphelion will be at $R_{DSM}$. Thus we have the following relationship:

$$R_p + R_{DSM} = 2a \qquad (30)$$

We know *a* from equation (29) and also $R_P$, thus we can calculate the radial distance required of the IP/DSM as follows:

$$R_{DSM} = 2a - R_p \qquad (31)$$

*Table 1 VLM for Earth Resonances (Aphelia)*

| Value of *n* | Approx. Duration of E-DSM-E, $nT_E$ (days) | Semi-major axis, *a* (au) | Placement of DSM - Aphelion (au) |
|---|---|---|---|
| 1 | 365 | 1.0 | 1.0 |
| 2 | 730 | 1.6 | 2.2 |
| 3 | 1096 | 2.1 | 3.2 |
| 4 | 1461 | 2.5 | 4.0 |

To demonstrate the application of equations (29) to (31), let us select planet P as Earth, thus the leveraging trajectory will be E-DSM-E. The leveraging resonances *n=2, n=3 & n=4* are provided in Table 1 and in Figure 7. Furthermore similar calculations can be performed but for P = Venus and the results are given in Table 2 and illustrated in Figure 8.

*Table 2 VLM for Venus Resonances (Aphelia)*

| Value of *n* | Approx. Duration of V-DSM-V, $nT_V$ (days) | Semi-major axis, *a* (au) | Placement of DSM - Aphelion (au) |
|---|---|---|---|
| 1 | 225 | 0.72 | 1.0 |
| 2 | 450 | 1.14 | 1.57 |
| 3 | 675 | 1.51 | 2.28 |
| 4 | 900 | 1.82 | 2.92 |

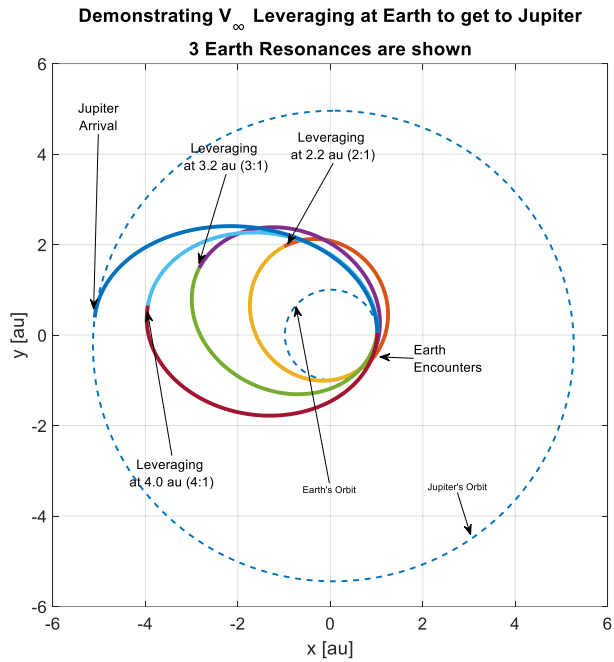

*Figure 7 Illustration of Earth Resonances for VLM*

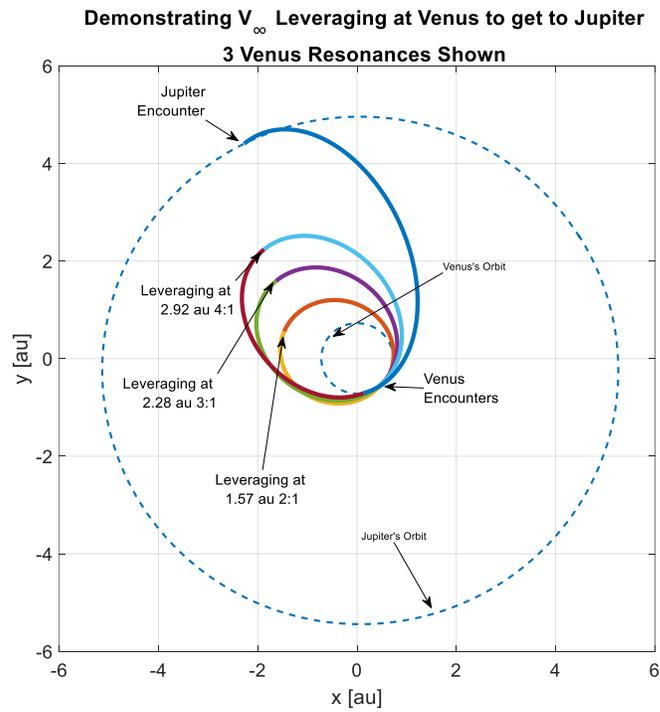

*Figure 8 Illustration of Venus Resonances for VLM*

### 4) Results of Investigations into ISO Missions which Use both SOMs and VLMs

The feasibility of missions to the ISO 1I/'Oumuamua was investigated through the 'Project Lyra' initiative. The KISS study [10] had already expounded a method of generating high heliocentric hyperbolic excess speeds into the interstellar medium (ISM), by utilising a SOM, and this was selected as a key trajectory option for Project Lyra research. Further in [10], an Earth VLM, with a 3 year resonance, was introduced to reduce the overall velocity increment $\Delta V$ and the magnitude of $C_3$ required at Earth by the selected launch system. Refer to Table 3 which provides results (with the IPs highlighted) of a solution by OITS for what shall be identified here as the Project Lyra Mission 1 scenario. Further analysis reveals that this $C_3$ and total $\Delta V$ increment combination can be achieved in practice, with significant payload masses, by a NASA Space Launch System.

As a consequence of the findings of the Interstellar Probe Concept Report [18], which detailed the results of a study into spacecraft missions to the ISM, Project Lyra shifted its attention from a Solar Oberth Manoeuvre, to a Jupiter Oberth Manoeuvre (JOM). This was because[18] effectively ditches the SOM in favour of either (a) a Passive Jupiter Gravitational Assist, or (b) a JOM.

However much one might argue the pros and cons of this decision, nevertheless it seemed pertinent to consider the JOM as an option for Project Lyra as it is does not have the additional challenges of a close approach to the sun (and consequent heat shield) and low technical readiness. Thus refer to Table 4 and Figure 10 for Project Lyra Mission 2 scenario. A VLM is inserted between the first return to Earth and the second, with a 2 year Earth resonance. Mission 2 is the same VEEGA sequence to Jupiter as [19].

*Table 3 Project Lyra Mission 1 with 2 Intermediate Points for VLM and SOM*

| Num | Encounter | Date UTC | Arrival Speed (kms$^{-1}$) | Departure Speed (kms$^{-1}$) | ΔV at Encounter (kms$^{-1}$) | Cumulative ΔV (kms$^{-1}$) | Periapsis Altitude (km) |
|---|---|---|---|---|---|---|---|
| 1 | Earth | 2030 JUN 09 | 0.0000 | 7.1062 | 7.1062 | 7.1062 | N/A |
| **2** | **IP @ 3.2au (VLM)** | **2031 NOV 25** | **11.5701** | **11.0052** | **0.6593** | **7.7656** | **N/A** |
| 3 | Earth | 2033 APR 17 | 11.9715 | 12.4288 | 0.3395 | 8.1051 | 200 |
| 4 | Jupiter | 2034 JUL 12 | 14.6104 | 14.4661 | 0.0674 | 8.1724 | 262748.9 |
| **5** | **IP @ 6 SR (SOM)** | **2036 FEB 24** | **251.2642** | **258.338** | **7.1710** | **15.3434** | **N/A** |
| 6 | Oumuamua | 2052 JUL 29 | 30.6801 | 30.6801 | 0.0000 | 15.3434 | N/A |

*Table 4 Project Lyra Mission 2 with 1 Intermediate Point for VLM*

| Num | Encounter | Date UTC | Arrival Speed (kms$^{-1}$) | Departure Speed (kms$^{-1}$) | ΔV at Encounter (kms$^{-1}$) | Cumulative ΔV (kms$^{-1}$) | Periapsis Altitude (km) |
|---|---|---|---|---|---|---|---|
| 1 | Earth | 2028 FEB 04 | 0.0000 | 4.2297 | 4.2297 | 4.2297 | N/A |
| 2 | Venus | 2028 JUN 23 | 6.7703 | 5.086 | 0.8464 | 5.0761 | 226.5 |
| 3 | Earth | 2029 FEB 27 | 7.5403 | 7.5403 | 0.0000 | 5.0761 | 2876.9 |
| **4** | **IP @ 2.2au (VLM)** | **2030 JAN 26** | **15.8353** | **15.5112** | **0.3464** | **5.4225** | **N/A** |
| 5 | Earth | 2031 FEB 11 | 9.5415 | 12.7324 | 2.2635 | 7.6860 | 200 |
| 6 | Jupiter | 2032 APR 16 | 16.9082 | 41.1406 | 10.4894 | 18.1754 | 200 |
| 7 | Oumuamua | 2050 JAN 29 | 22.4985 | 22.4985 | 0.0000 | 18.1754 | N/A |

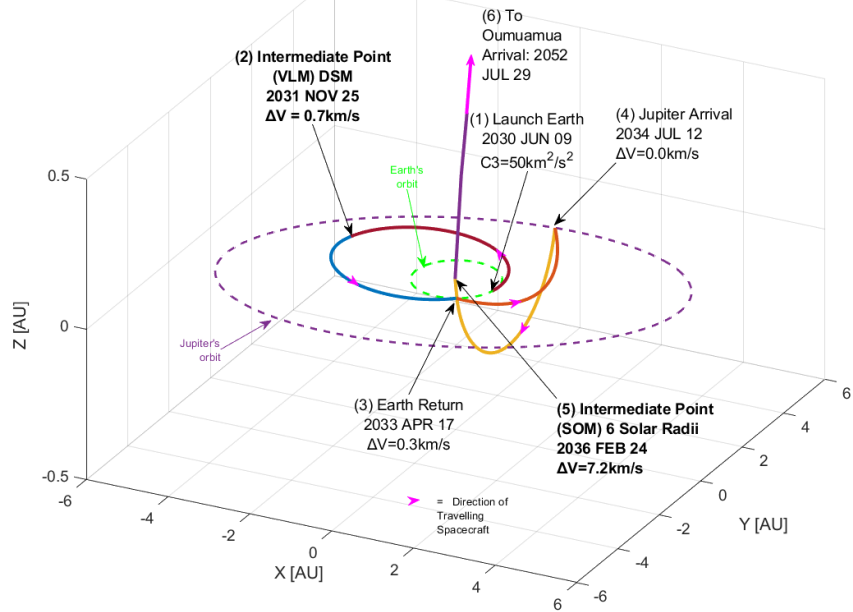

*Figure 9 Project Lyra Mission 1*

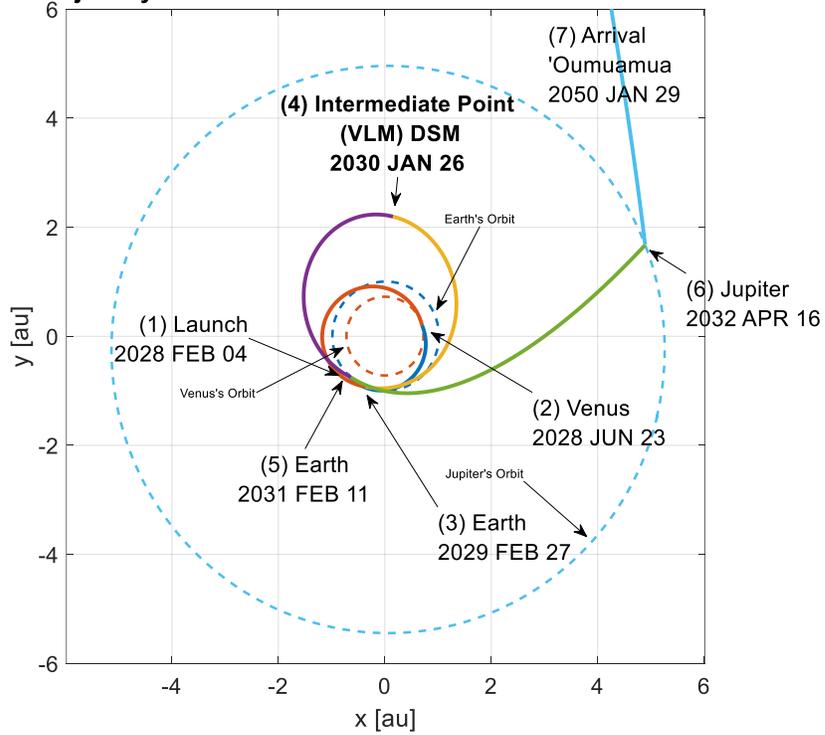

*Figure 10 Project Lyra Mission 2*

## 5) Conclusions

The simplicity of the concept of the Intermediate Point belies its flexibility and utility, allowing solutions of interplanetary missions with specified $V_\infty$ Leveraging Manoeuvres for any chosen planet (in particular, Earth and Venus), as long as the corresponding calculation of aphelion distance for the Deep Space Manoeuvre is performed. Solar Oberth Manoeuvres can also be modelled using IPs, and rapid optimal trajectory solutions to ISOs can be generated allowing in-depth analysis of mission feasibility. A feature not expounded in this paper is the capability of studying missions to distant 'virtual' objects belonging to our solar system, where the heliocentric distance to the object can be specified by an IP, and in addition the heliocentric longitude and latitude can be set as fixed parameters.